\documentclass[11pt,twoside]{article}  
\usepackage{apn3conf}

\begin{document}   
\def \dz{{\mathrm{d}}z}
\def \jcp{J. Comput. Phys.}
\def \ea{{\it et al.}}
\def \etal{{\it et al.}}
\def \curl{{\nabla\times}}
\def \eg{{\it e.g.}}
\def \ie{{\it i.e.}}
\def \etc{{\it etc.}}
\def\lsim{\mathrel{\rlap{\lower4pt\hbox{\hskip1pt$\sim$}}
    \raise1pt\hbox{$<$}}}
\def\gsim{\mathrel{\rlap{\lower4pt\hbox{\hskip1pt$\sim$}}
    \raise1pt\hbox{$>$}}}
\def \msol{\rm{M}$_\odot$}
\def \lsol{\rm{L}$_\odot$}
\def \mdot{\rm{M}$_\odot$~yr$^{-1}$}
\def \rsol{\rm{R}$_\odot$}
\def \kms{km~$\rm{s}^{-1}$}
\def \kmsbig{$^{\textrm{km}}/_{\textrm{s}}$}
\def \kmssmall{$^{\mathrm{km}}/_{\mathrm{s}}$}
\def \cc{$\rm{cm}^{-3}$ }
\def \arcs{\char'175}
\def \lam{$\lambda$}
\def \micron{$\mu$m{}}
\newcommand{\sfrac}[2]{\,{}^{#1}\!/_{#2}}
\newcommand{\beq}{\begin{equation}}
\newcommand{\eeq}{\end{equation}}
\newcommand{\bdm}{\begin{displaymath}}
\newcommand{\edm}{\end{displaymath}}
%
%
%
%

\title{MHD Disk Winds in PNe and pPNe}

%

\author{Adam Frank}
\affil{Department of Physics and Astronomy Department, University,
of Rochester, Rochester, NY, 14610, \\
afrank@pas.rochester.edu}


\contact{Adam Frank} \email{afrank@pas.rochester.edu}

%
%
%
%
%

\paindex{Frank, A.} \aindex{Headroom, M.} \aindex{Duck, D.}

%
%

\authormark{Frank}

%

\keywords{magnetic fields, accretion disks, jets}


\begin{abstract}          
Winds from accretion disks have been proposed as the driving
source for precessing jets and extreme bipolar morphologies in
Planetary Nebulae (PNe) and proto-PNe (pPNe). Here we apply MHD
disk wind models to PNe and pPNe by estimating separately the
asymptotic MHD wind velocities and mass loss rates. We show that
the resulting winds can recover the observed momentum and energy
input rates for PNe and pPNe.
\end{abstract}



\noindent{\bf Introduction} While considerable progress has been
made in understanding the hydrodynamic shaping of elliptical and
bilobed PNe (Balick \& Frank 2002), the origin of extreme
butterfly nebulae as well as jets in PNe continues to pose a
number of problems for theorists.

In particular a formidable problem for pPNe concerns the total
momentum and energy in the outflows. A number of observational
studies have shown that radiatively accelerated winds in pPNe
cannot account for the high momentum and energy implied by CO
profiles (Bujarrabal \ea 2001). Thus the total flow momentum $\Pi$
is such that $\Pi >> (L_*/c)\Delta t$, where L$_*$ is the stellar
luminosity emitted during the pPNe outflow expansion lifetime
$\Delta t$. In light of these results both the {\it launching and
collimation} of winds in pPNe becomes problematic.  Note that
while both the dominant hydrodynamic theory for shaping PNe, (the
Generalized Interacting Stellar Winds model: Balick 1987; Icke
1988), and models invoking an initially weak toroidal magnetic
field (Garcia-Segura \ea 1999) can produce jets neither can
account for the momentum excesses in pPNe.

Thus there remains considerable uncertainty about the processes
which produce collimated jets/outflows in pPNe and PNe. Other
systems which produce jets such as YSOs, AGN and micro-quasars
have been modelled via a combination of magnetic and centrifugal
forces from accretion disks (Konigl \& Pudritz 2000).  The success
of these {\it Magneto-centrifugal Launching} (MCL) models is such
that it is worthwhile considering if such a scenario can be
applied to PNe and pPNe.  While models such as Morris 1987, Soker
\& Livio 1994 and Soker \& Rapport 2001 have relied heavily on
collimated winds from disks these works did not specify how such
winds are launched or collimated. Thus application of MCL disk
wind models to PNe and pPNe would close an important gap in
building a new MHD disk wind paradigm for these systems (Blackman,
Welch \& Frank (2001): BFW01, Blackman \ea ~2001).

Here we summarize new calculations from Frank \& Blackman 2004 who
derive scaling relations from the equations for MCL and separately
estimate MHD disk wind mass outflow rates and asymptotic outflow
velocities. These relations are then applied to PNe and pPNe. In
these calculations a disk dynamo is invoked to produce the
requisite fields.

\bigskip
\noindent {\bf Disk Accretion Rate in PNe:} In order to produce a
more detailed comparison of MHD disk winds with PNe it is
necessary to have a model for PNe accretion disks. It is unlikely
that an accretion disk could survive the long main sequence
lifetime of a PN central star. Thus, unlike YSOs and AGN,
accretion disks in PNe systems must form via binary interactions.
Disks may form around secondaries via Roche lobe overflow or
accretion of the dense AGB wind (Mastrodemos \& Morris 1998). Such
systems would be similar to symbiotic stars.

Mastrodemos \& Morris (1998) found steady accretion disks could
form around a white dwarf companion orbiting a AGB star with
$\dot{M}_{agb} \approx 10^{-5}$ \mdot.  The ratio they found in
their models of $\dot{M}_d/\dot{M}_{agb} \approx .05 ~-~ .005$ is
consistent with expectations from basic theory.

Accretion disks could also form around the primary after CE
evolution and disruption of the secondary star (Reyes-Ruiz \&
Lopez 1999: RRL99).  This model implies a finite lifetime for the
disk as the mass reservoir of the disrupted companion is slowly
drained onto the primary.  A description of disk formation in PNe
has been given in RRL99 who found that systems with a primary
consisting of an evolved AGB star with mass $M_* \approx 2.6 ~-~
3.6 ~M_\odot$, a low mass secondary ($\le 0.08$ \msol) and an
initial binary separation of $< 200 R_\odot$ may produce disks.

RRL99 also found the disk accretion rate to evolve in time with in
a power-law manner. $\dot{M}_d  =  \dot{M}_{do} \left( \frac{t}{1
~yr} \right)^{-5/4} {\rm{M}_\odot~yr^{-1}} $ Typical values of the
scale is $\dot{M}_{do} = 10^{-3}$ \mdot.

\bigskip
\noindent {\bf Mass Outflow Rate and Wind Speed from MCL Theory:}
The basic physics of magneto-centrifugal launching of winds and
jets is well studied when a magnetic field distribution is imposed
on the disk. In Frank \& Blackman 2004 it was shown how to combine
Poynting flux driven outflows with asymptotic wind solutions and
mean field dynamo theory to estimate the asymptotic wind speed and
the outflow accretion rates.

Magneto-centrifugal launching is a means of converting
gravitational binding energy in an accreting source into kinetic
energy of an outflowing wind.  The magnetic fields act as a drive
belt to extract angular momentum from the anchoring rotator and
launch the wind.  The magnetic luminosity, or equivalently, the
maximum magnetic power available for a wind can be obtained from
the integrated  Poynting flux (BFW01).
\begin{eqnarray}
L_w & = & \frac{1}{2} {\dot M}_w u_{\infty}^2 \sim L_{mag} \\
    & \equiv &
\int ({\bf E} \times {\bf B}) \cdot d{\bf S}_A \sim
\int_{r_{i}}^{r_{A}(r_i)} \Omega (r)r B_pB_\phi rdr \sim  B_A^2
\Omega_0 r_A^3, \label{magicF}
\end{eqnarray}
where $r_0$ is the disk inner radius and $B_A= B_\phi \sim B_p$ at
the Alfv\'en surface

More work is required to estimate the mass outflow rate and
outflow speed separately. The MCL problem requires the
construction of solutions for a steady, ideal, isothermal
magnetohydrodynamic flow. The isothermal assumption eliminates the
need for solving the energy equation, but more complex assumptions
can be used, \eg ~a polytropic law.  Blackman \& Frank 2004 began
with an expression for the magnetic Bernoulli constant which
connects the wind at infinity to that at the footpoints of the
disk
\begin{equation} \frac{1}{2}(u_p^2 +
\Omega^2 r^2) + \Phi + w + \Omega_o(\Omega_o r_A^2- \Omega r^2) =
U(a) = const(a) \label{berneq}.
\end{equation}
\noindent where $u_p$ is the polodial speed at infinity, $r_A$ is
the Alfven radius where $u_p = u_A$.  This equation can then be
combined with the momentum equation for the flow with an effective
potential for a cold plasma parcel tied to a rotating field line
(Blandford \& Payne (1982))

\beq
{\bf u}\cdot\nabla u_r = -\partial_r \Phi_{eff} = \left({GM_*
\over r_0}\right)\left({r \over r_0^2}-{rr_0\over (z^2 +
r^2)^{3/2}}\right).
\label{bp}
\eeq.

\noindent Beginning with these two equations one can estimate the
properties of the wind at large distances from the disk source if
the field strength in the disk can be derived.

Magnetic fields may form in these disks via dynamo processes. The
topology of such a field, (the ratio of poloidal $B_p$ and
toroidal $B_\phi$ in the disk) and its subsequent value in the
coronae remains a subject of considerable discourse.  Frank \&
Blackman 2004 assumed a dynamo driven field in the disk that led
to a primarily polodial field in a magneto-hydrostatic disk corona
where the wind launches.  From the expressions derived for the
disk field along with those estimated from the governing MHD
equations Frank \& Frank 2004 provided estimates for the wind
properties. We quote these below,

\beq {\dot M}_w\sim 0.1 \alpha_{ss}{r_0\over h_0}{\dot M}_d,
\label{mdot5} \eeq

\beq u_\infty \sim 2.1\Omega_0 r_0. \label{uinf3} \eeq

\beq L_{mag}\simeq 0.22 \alpha_{ss}{r_0\over h_0}{\dot M}_d
\Omega_0^2 r_0^2, \label{maglum} \eeq

\beq \dot{\Pi} \sim L_{mag}/u_{\infty} \simeq 0.14
\alpha_{ss}{r_0\over h_0}{\dot M}_d \Omega_0 r_0. \label{momentum}
\eeq

where $\alpha_{ss}$ is a dimensionless parameter associated with
the disk viscosity and $h_0$ is the disk scale height.  The third
and fourth expressions give estimates of the rate in which energy
and momentum are input by the MCL disk wind into the ambient
medium.


\bigskip \noindent {\bf Disk Winds Models for PNe and pPNe:}
For "classic" PNe, a total mass of $M_{pn} \approx .1$
\msol ~must be accelerated to velocities of $u_{pn} \approx 40
~km/s$ in a timescale of order $\Delta t_{pn} \approx 10000 ~y$.
This gives $\dot{\Pi} = M_{pn} u_{pn}/\Delta t_{pn} \approx
10^{27} ~g ~ cm/s^{2}$ and $\dot{E} = M_{pn} u_{pn}^2/\Delta
t_{pn} \approx 10^{34}~erg/s$.

For pPNe Bujarrabal \ea (2001) found high total outflow momentum
$10^{36} < \Pi/(g ~cm ~s^{-1}) < 10^{40}$ and total outflow energy
$10^{41} < E/(erg ~s^{-1}) < 10^{47}$. These values can be
converted into momentum and energy injection rates using an
assumed injection or "acceleration" timescale $\Delta t$,
$\dot{\Pi} = \Pi/\Delta t, ~L = E/\Delta t$.

The question which arises is: can these energy and momentum
budgets be met with disk wind models. In what follows we use the
relations from Frank \& Blackman 2004 quoted above and assume that
the inner edge of the disk extends to the stellar surface and use
$r_0 = r_i = r_*$.

\noindent {\bf PNe Solutions:}  In this case we assume that the
star which produces the jet is a proto-WD with an AGB companion
(Soker \& Rappaport 2001). Thus accretion rates of $\dot{M}_d
\approx 10^{-6}$ \msol $y^{-1}$ are reasonable. To evaluate the
expressions above we choose $M_s = 0.6$ \msol and a disk with
$\alpha_{ss} = 0.1$ and $r_0/h_0 = 10$.

For PNe central star parameters ($T_* = 10^5 ~K$, $L_* = 5000
~L_\odot$ such that $r_i = 1.64 \times 10^{10} ~cm$) we find the
following conditions for the wind from the equations above
\begin{eqnarray}
\dot{M_w} & = & 1\times10^{-7} ~\rm{M}_\odot~yr^{-1}
\left(\frac{\dot{M}_a} {10^{-6}
~\rm{M}_\odot~yr^{-1}}\right) \\
u_\infty & = & 1.25\times10^3 ~km/s
\left(\frac{M_*}{.6~\rm{M}_\odot}\right)^{1/2}
\left(\frac{r_0}{.23 ~\rm{R}_\odot}\right)^{-1/2}. \\
\end{eqnarray}
\noindent Thus using typical conditions for PNe central stars, the
scaling relations derived from the MHD equations yield disk wind
parameters well matched with observations.

\noindent {\bf pPNe Solutions:} While the mass loss rates and
velocities are known for PNe winds the situation for pPNe is not
as clear. In general what is observed in pPNe is the total mass in
the outflows.  Velocities are also uncertain as only properties of
swept-up material may be directly determined.

We assume a post-AGB star with mass $M_s = 0.6$ \msol, $T_* =
10,000 ~K$ and $L_* = 5 \times 10^3$ \lsol ~ which, assuming a
blackbody, yields a radius of $r_* = 1.6 \times 10^{12} ~cm~= 23$
\rsol. Note that such a star has an escape velocity of $u_{esc} =
98 ~km ~s^{-1}$

Achieving the high momentum input rates observed in pPNe via MCL
disk wind models will necessitate high accretion rates. We use an
accretion rate of $\dot{M}_d = 1 \times 10^{-4}$ \mdot which is
the $200$ year average of that found by RRL99 for their case A
Once again we choose a disk with $\alpha_{ss} = 0.1$ and $r_0/h_0
= 10$.

Assuming $r_0 = r_*$ along with the other parameter values given
above, the key wind quantities $\dot{M}_w$, $u_w$, $L_m = L_w,
\dot{E}$ are,
\begin{eqnarray}
\dot{M_w} & = & 1. \times10^{-5} ~~\rm{M}_\odot~yr^{-1}~
\left(\frac{\dot{M}_d}{10^{-4} ~\rm{M}_\odot~yr^{-1}}\right) \\
u_\infty & \simeq & 146 ~km/s \left(\frac{M_*}{.6 ~\rm{M}_\odot
}\right)^{1/2}\left(\frac{R_i}{23 ~\rm{R}_\odot}\right)^{-1/2} \\
L_m & \simeq & 6.7 \times 10^{34} erg ~s^{-1}
\left(\frac{\dot{M}_d}{10^{-4} ~\rm{M}_\odot~yr^{-1}}\right)
\left(\frac{M_*}{.6 ~\rm{M}_\odot }\right) \left(\frac{R_i}{23
~\rm{R}_\odot}\right)^{-1}
\end{eqnarray}
Note that $L_w \Delta t \approx 10^{44}$, a value in the middle of
the range found by Bujarrabal \ea 2001.  Note also that the
solution above has $u_\infty \approx 1.5 u_{esc}$.  Since
$u_\infty \approx u_{esc}$ the higher velocity outflows seen in
some pPNe would require more disks around more compact central
sources.

Given a model for the temporal history of the disk accretion, the
total energy and momentum for the outflows can be found. Replacing
$\dot{M}_d$ with $\dot{M}_d(t)$ from RRL99 and integrating gives
\begin{eqnarray}
E    & = & \int \frac{1}{2}\dot{M}_w u_\infty^2 dt \approx 1.3
\times10^{44} ~erg
~\left[1 - \left(\frac{1 ~yr}{t}\right)^{1/4})\right] \\
\Pi  & = & \int \dot{M}_w u_\infty dt \approx 1.8\times10^{37} ~g
~cm ~s^{-1}~ \left[1 - \left(\frac{1 ~yr}{t}\right)^{1/4}\right]
\end{eqnarray}
These results show that the MCL disk wind models can achieve both
energy and momentum injection rates {\it as well as} the total
energy and momentum required to account for many pPNe described by
Bujarrabal et al (2001). The total energy and momenta budgets we
find from these solutions fall well within the range of pPNe
outflows with momentum excesses. Taken together with our previous
calculations for "classic" PNe winds, these results confirm the
predictions of BFW01 that magnetized disk winds can account for
much of the outflow phenomena associated collimated outflows in
the late stage of stellar evolution.

The results above indicate that collimated flows which form from
transient disks in the pPNe stage will appear as dense knots in
mature PNe flows. This may also serve to explain the presence of
so-called FLIERS (Fast Low Ionization Emission Regions) seen in
some PNe. The mass loss rate in the winds derived above rapidly
decrease with time. Thus the bulk of the jet's mass will lie near
its head. As the material in the disk is accreted onto the star
the jet will eventually shut-off leaving the dense knot to
continue its propagation through the surrounding slow wind.

When the star makes its transition to a hot central star of a PNe
its fast, tenuous spherical wind sweeps up a shell of the slow AGB
wind material.  The shell's expansion speed will typically be of
order $40 ~km/s$ and it will not catch up to the head of the jet.
Thus during the PNe phase the jet head will appear as a dense,
fast moving knot which should lie outside the PNe wind blown
bubble. We note that masses of FLIERs are estimated to be of order
$10^{-4} - 10^{-5} ~\rm{M}_\odot$ which is reasonable for the
models presented above. FLIER velocities can be lower than the
$\approx 100 ~km/s$ calculated above but deceleration of the jet
head will occur via interaction with the environment. We note also
that hydrodynamic simulations of PNe jets in which the jet ram
pressure decreases in time (as would occur for our model) show
characteristic patterns of backward pointing bow-shocks (apex
pointing back towards to the star).   If such results are robust,
the jets produced by disk winds in our scenario above may also
yield similar morphologies.

\bigskip\noindent {\bf Discussion and Conclusions} Our results for
pPNe show that momentum excesses need not occur for outflows
driven by MCL winds. While this is encouraging in terms of finding
a mechanism for driving pPNe outflows the solutions require fairly
high accretion rates ($ > 10^{-5}$ \mdot).  It is not clear if
such conditions can be achieved with the frequency required by
observations. While solutions of RRL99 yield accretion rates and
time dependencies which lead to the correct outflow momenta and
energetics, their models place fairly stringent limitations on the
nature of the binaries that form disks from disrupted companions.
If accretion onto undetected compact orbiting companions is
invoked Soker \& Rappaport (2001) then higher values of
$\dot{M}_d$ may not be required.

We note that a robust prediction of our models is the ratio of
wind mass loss rate to accretion rate, i.e.
$\dot{M}_{w}/\dot{M}_{a} \approx .1$. This is true for most MCL
disk wind models and can be seen as a target prediction which can
be explored observationally.

This paper comprises a step beyond Blackman Frank Welch (2001) in
establishing the efficacy of MHD paradigms for pPNe/PNe in which
strong magnetic fields play a role in both launching and
collimating the flows.  It is also worth noting that recent
simulation results by Matt, Blackman \& Frank 2004 confirm that
the exposed rapidly rotating magnetic core model can produce well
collimated outflows.

{\bf ACKNOWLEDGMENTS} Support by NSF grants AST-9702484,
AST-0098442, NASA grant NAG5-8428, HST grant,  DOE grant
DE-FG02-00ER54600. and the Laboratory for Laser Energetics.

%
%
%


\end{document}